# Orientation-Driven Large Magnetic Hysteresis of Er(III) Cyclooctatetraenide-Based Single-Ion Magnets Adsorbed on Ag(100)

*Vladyslav Romankov, Moritz Bernhardt, Martin Heinrich, Diana Vaclavkova, Katie Harriman, Niéli Daffé, Bernard Delley, Maciej Damian Korzyński, Matthias Muntwiler, Christophe Copéret, Muralee Murugesu, Frithjof Nolting, Jan Dreiser\**

V. Romankov, M. Heinrich, D. Vaclavkova, N. Daffé, B. Delley, M. Muntwiler, F. Nolting, J. Dreiser
Paul Scherrer Institut, Forschungsstrasse 111, CH-5232, Villigen PSI, Switzerland
E-mail: jan.dreiser@psi.ch

V. Romankov, M. Heinrich
Swiss Nanoscience Institute, University of Basel, Klingelbergstrasse 82, 4056 Basel, Switzerland

K. Harriman, M. Murugesu
Department of Chemistry and Biomolecular Sciences, University of Ottawa, 150 Louis-Pasteur Pvt, ONK1N6N5, Ottawa, Canada.

M. Bernhardt, C. Copéret
Department of Inorganic Chemistry and Applied Biosciences, ETH Zürich, Vladimir-Prelog-Weg 1-5/10, 8093 Zürich, Switzerland

M. D. Korzyński,
Department of Chemical and Physical Sciences, University of Toronto Mississauga, 3359 Mississauga Road, ON, L5L 1C6, Mississauga, Canada

**The molecular self-assembly and the magnetic properties of two cyclooctatetraenide (COT) - based single-ion magnets (SIM) adsorbed on Ag(100) in the sub-monolayer range are reported. Our study combines scanning-tunneling microscopy, X-ray photoemission spectroscopy and polarized X-ray absorption spectroscopy to show that Cp\*ErCOT (Cp\* = 1,2,3,4,5-pentamethylcyclopentadienide anion) SIMs self-assemble as alternating compact parallel rows including standing-up and lying-down conformations, following the main crystallographic directions of the substrate. Conversely, K[Er(COT)$_2$], obtained from subliming the [K(18-c-6)][Er(COT)$_2$]·2THF salt, forms uniaxially ordered domains with the (COT)$^{2-}$ rings perpendicular to the substrate plane. The polarization-dependent X-ray absorption spectra reproduced by the multiX simulations suggest that the strong in-plane magnetic anisotropy of K[Er(COT)$_2$]/Ag(100) and the weak out-of-plane anisotropy of Cp\*ErCOT/Ag(100) can be attributed to the strikingly different surface ordering of these two complexes. Compared to the bulk phase, surface-supported K[Er(COT)$_2$] exhibits a similarly large hysteresis opening, while the Cp\*ErCOT shows a rather small opening. This result reveals that despite structural similarities, the two organometallic SMMs have strongly different magnetic properties when adsorbed on the metal substrate, attributed to the different orientations and the resulting interactions of the ligand rings with the surface.**

1. Introduction

Single-molecule magnets (SMMs) exhibit magnetic hysteresis without the presence of magnetic long-range order. This group of materials includes exchange-coupled molecular clusters with several metal ions and molecular complexes based on only one ion.[1–8] The latter are often denoted as single-ion magnets (SIM). In this context, lanthanide ions are appealing due to their large unquenched orbital magnetic moments, high spin ground states and the possibility of tuning the magnetic anisotropy via proper ligand environment.[9,10] Sandwich-type complexes based on phthalocyanine, cyclopentadienide (Cp) and/or cyclooctatetraenide (COT) ligands, as well as their derivatives, have been shown to induce large magnetic anisotropy on lanthanide ions such as Tb$^{3+}$, Dy$^{3+}$ and Er$^{3+}$.[5,11–16] In some cases it has been shown that the complexes preserve the hysteresis opening even above liquid nitrogen temperature.[17] This makes these coordination compounds attractive for ultra-high-density memory applications. In turn, for any device applications, the two-dimensional ordering of the SIMs on planar surfaces is of vital importance to access the magnetic information by local probes, for example using a scanning probe microscopy tip.[18] The molecule-substrate interaction is an important factor affecting most often negatively the hysteresis loop openings upon surface adsorption of SMMs and SIMs.[18–24] In other cases, strong molecule-surface interaction promotes charge transfer that can completely alter the magnetic properties of molecules on metals in the monolayer range.[25] A few exceptions rely on the use of long chemical linkers or stiff buffer layers to quench vibrational relaxation pathways, while also eliminating the hybridization with the metallic surface.[26–28]

Here we study two surface-adsorbed mononuclear $Er^{3+}$ sandwich-type SIMs, the neutral Cp*ErCOT (with (Cp*)$^-$ the 1,2,3,4,5-pentamethylcyclopentadienide anion and (COT)$^{2-}$ the cyclooctatetraenide dianion)[29] and the ionic K[Er(COT)$_2$], with the structures shown in **Figure 1**a. K[Er(COT)$_2$] is obtained by sublimation of [K(18-c-6)][Er(COT)$_2$]·2THF (where THF is tetrahydrofuran and "18-c-6" denotes the 18-crown-6 ether). Both Cp*ErCOT and [K(18-c-6)][Er(COT)$_2$]·2THF show strong magnetic anisotropy with the easy axis aligned perpendicular to the almost parallel ligand rings,[30–33] similar to other COT-based (and their functionalized counterparts) SMMs.[34–38]

We focus on comparing the self-assembly and the magnetic properties of these SIMs in the sub-monolayer (ML) to the few-layer range adsorbed on the (100) surface of an Ag single crystal. A strikingly different ordering and self-assembly of the two compounds is unveiled by scanning tunneling microscopy (STM) and X-ray linear dichroism (XLD). Point-charge-based simulations as implemented in the multiX code[39] are compared to polarized X-ray absorption spectra to understand the net magnetic anisotropy in connection with the ordering of the SIMs. Finally, X-ray magnetic circular dichroism (XMCD) is used to detect the element-specific magnetic moments and the magnetic hysteresis loops.

## 2. Results and discussion
### 2.1. Chemical characterization

Samples of surface-adsorbed K[Er(COT)$_2$] and Cp*ErCOT were obtained by sublimation of polycrystalline powder onto a freshly prepared Ag(100) surface in ultra-high vacuum. A ML corresponds to a densely packed single layer of molecules. Normalized XPS spectra of K[Er(COT)$_2$](~0.5 ML )/Ag(100) and Cp*ErCOT(~1 ML)/Ag(100) are reported in Figure 1b and 1c, respectively. The K[Er(COT)$_2$]/Ag(100) has the main peak of the C 1s core level at a binding energy of 283.9 eV (FWHM 0.8 eV) and a smaller peak at 286.9 eV (FWHM 1.1 eV).

We ascribe the former peak to the sp$^2$-hybridized carbon atoms of (COT)$^{2-}$, while the small feature at higher binding energy is consistent with the presence of oxygen-bound carbon. Since the [K(18-c-6)][Er(COT)$_2$]·2THF salt was used as source material it is not surprising to find a small fraction of THF and (18-c-6) ether molecules deposited on the surface as well.[40,41] However, as shown by XPS spectra recorded at different crucible temperatures, the volatile crown ether and THF are almost completely removed from the source material below the sublimation temperature of K[Er(COT)$_2$] (see Supplementary Information (SI)). The lower binding energy of the C 1s levels with respect to reported charge neutral metallocenes[42] is attributed to the charge separation and the resulting coverage dependent electric dipole formation that the complex undergoes upon direct interaction with the substrate, as discussed further below and in the SI. Note that because of this the C 1s binding energies cannot be directly compared with the ones of Cp*ErCOT and literature references. Nevertheless, the atomic ratio of Er, K and sp$^2$ C corresponds to the stoichiometry expected for K[Er(COT)$_2$] complexes (see discussion in the SI).

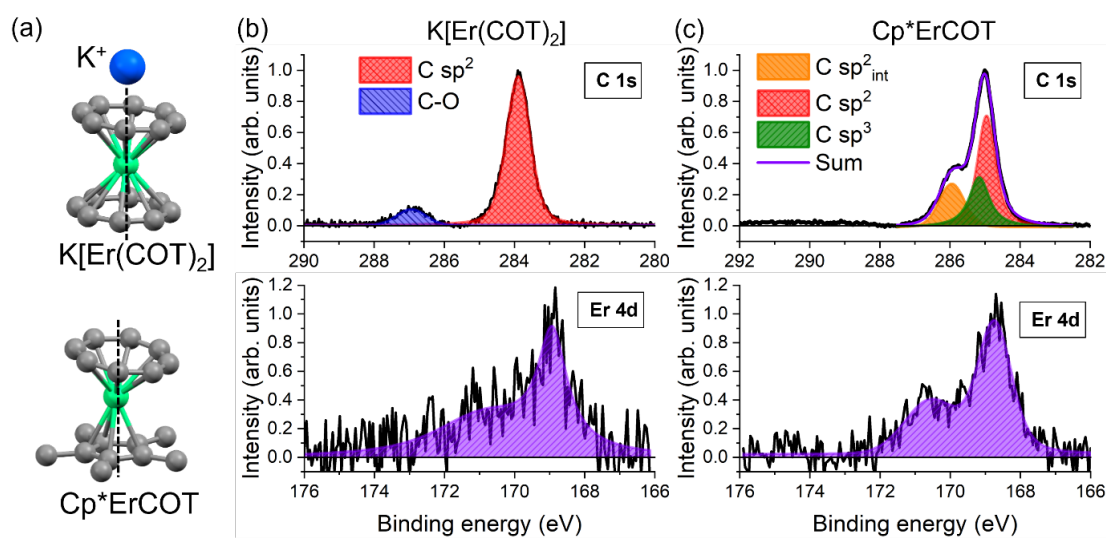

**Figure 1**. (a) Molecular structures of K[Er(COT)$_2$] (top) and Cp*ErCOT (bottom). Color code: green: Er$^{3+}$; grey: C; blue: K. Hydrogens are omitted for simplicity. C 1s and Er 4d normalized core level spectra of (b) K[Er(COT)$_2$](~0.5 ML)/Ag(100) and (c) Cp*ErCOT(~1 ML)/Ag(100).

The analysis of the C 1s core level spectrum of Cp*ErCOT/Ag(100) as shown in Fig. 1c yields three contributions: a peak at 284.9 eV (FWHM 0.6 eV), attributed to the sp$^2$-hybridized carbon atoms of the aromatic rings;[42–44] a peak at 285.2 eV (FWHM 0.7 eV), attributed to the sp$^3$ carbon atoms of the methyl groups in the Cp* ligand;[45] and a peak at 285.9 eV (FWHM 0.8 eV), attributed to the sp$^2$-hybridized carbon atoms of the COT ligands interacting with the metal substrate, as discussed further below. The Er 4d core levels appear at characteristic energies for the trivalent Er$^{3+}$ ion,[46] with the main peak of the K[Er(COT)$_2$] centered at 168.9 eV, while the one of Cp*ErCOT SMMs lies at 168.7 eV. In order to obtain the stoichiometry from the XPS scans, the energy-dependent cross-sections and asymmetry parameters were taken from tables reported in the literature.[47] The extracted C:Er atomic ratio for the Cp*ErCOT SMMs is 21:1, while the K[Er(COT)$_2$] shows a C:Er ratio of 16:1, a K:Er ratio of 0.8:1 and O:Er ratio is 2.5:1. Within the error bar of ~30%, which is mainly due to the weak Er 4d signal used to estimate the ratios, these values are in excellent agreement with the atomic ratios expected from the molecular structures, pointing to the integrity of the complexes on the surface.

**2.2 Surface ordering**

The STM images of Cp*ErCOT(~1 ML)/Ag(100) are presented in **Figure 2**a – 2c. Alternating rows of brighter and darker spots can be identified in the 20 × 20 nm$^2$ image in Figure 2a. The rows are oriented along the [010] and [001] crystallographic directions of the Ag substrate and they form domains of a few tens of nanometers, that cover densely the surface. The average distance between two rows of bright spots amounts to 1.48 ± 0.04 nm, while the distance between two height maxima within the same row is 0.84 ± 0.05 nm. We ascribe the different spots to two different adsorption conformations, with the orientation of the complexes identified by their main rotational axis (dashed lines in Figure 1a). In Figure 2a the distances between the spots are consistent with alternating rows of standing-up and lying-down complexes on the surface, as detailed further below. In some parts of the islands, the ordering of the complexes follows a herringbone structure as shown by the black pattern in Figure 2a. This can be explained by the orientation of the main molecular axes of the lying-down complexes being rotated at an angle in the surface plane, (see overlay in Figure 2c). The distance between two standing-up complexes in this rotated direction amounts to $d = 1.54 ± 0.04$ nm, perfectly matching the distance of 1.55 nm (at 10 K) that the complexes have along the [-101] direction in the crystal structure of pure Cp*ErCOT, with the same standing-up and lying-down geometry.[29] However, the surface ordering does not correspond to the one in any of the planes present in the crystal structure of the parent compound.

The bright spots in Figure 2b are ascribed to the up-right complexes, with the (Cp*)$^-$ ring being on the top, resembling in size and shape the one reported for [Cp*Ru]$^+$ ions on graphene.[48] The bright lobes of the lying-down complexes are attributed to the methyl groups, while the darker areas are attributed to the (COT)$^-$ rings, poorly conducting in the direction perpendicular to the π-bonds.[49,50] From Figure 2b and 2c, the molecular density is estimated to ~1.7 molecules/nm$^2$.

Figure 2c shows a zoom into part of the area reported in Figure 2a, sampled at a bias voltage of $V_b = 2$ V and a current setpoint of $I_s = 500$ pA. In this condition, the apparent heights of the two rows become comparable and the lying-down complexes are seen as single bright spots, possibly due to an enhanced conduction through the $Er^{3+}$ ion at these imaging conditions. The overlay presented in the Figure shows that the herringbone structure mentioned previously probably originates from the alternating orientation of the complexes in the lying-down rows. The superstructure formed by the complex can be described by the unit cell reported in Figure 2c, with $a_1 \times a_2 = 0.84 \times 2.96$ nm$^2$ and the vectors oriented along the [010] and [001] directions of the substrate. The black dotted line shows the lateral shift of $0.15 \pm 0.02$ nm of every second row. Note that defect rows are present in form of three consecutive rows of standing-up complexes as seen, e.g., on the upper right of Figure 2a. Further STM images showing the coverage and the terrace steps are reported in the SI.

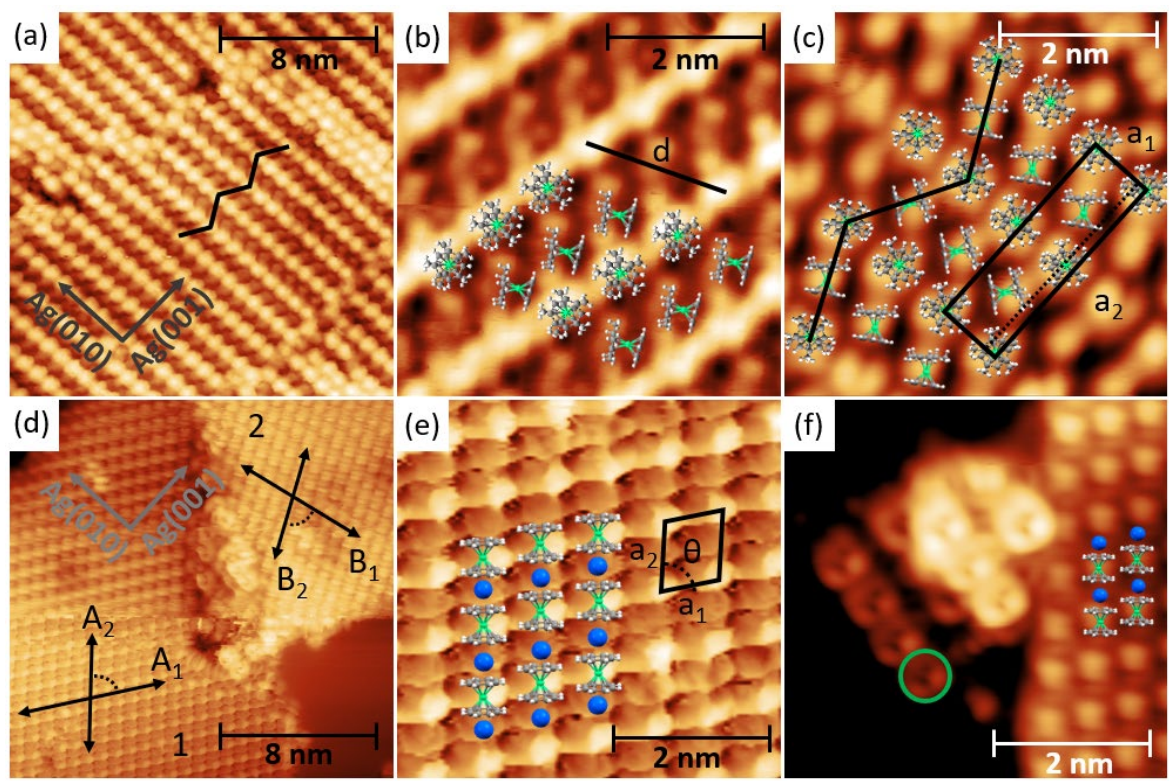

**Figure 2**. Constant-current STM images at 4.5 K of (a-c) Cp*ErCOT and (d-f) K[Er(COT)$_2$] adsorbed on Ag(100). Imaging conditions: (a) 0.8 V; 200 pA (b) 0.25 V; 50 pA; (c) 2 V; 500 pA; (d) -2 V; 50 pA; (e) zoom into (d); (f) -2 V; 50 pA. Molecular overlays, the herringbone structure, the molecular orientation indicated by the black arrows and the unit cells are shown, as explained in the main text. Color code: green: Er; grey: C; blue: K; white: H.

The STM images of K[Er(COT)$_2$](0.5 ML)/Ag(100) are reported in Figure 2d – 2f. Figure 2d shows two different domains of highly ordered complexes, labeled as 1 and 2, with the orientation of the rows indicated by black arrows. In both domains, the directions labeled $A_1$, $A_2$ and $B_1$, $B_2$ form an angle of 75°. With respect to the Ag(001), the direction labeled $A_1$ is rotated clockwise by ~35°, while the one labeled $B_1$ is rotated clockwise by ~80°. The complexes of both domains form highly oriented compact rows, as seen in the zoom into domain 1, reported in Figure 2e. The latter Figure shows a repeating pattern of ovals with an approximate size of 0.44 nm × 0.74 nm exhibiting a rhombic assembly, with a brighter feature on every second oval along the $A_2$ direction of the domain ($B_2$ on domain 2). The size of the ovals is consistent with the one of the [Er(COT)$_2$]$^-$ anions lying-down on the substrate (~0.4 × 0.6 nm),[29] with the molecular axis parallel to the surface plane. The modulation of the local density of states is likely given by the alternation of K$^+$ and Er$^{3+}$ ions, separated by standing COT$^{2-}$ rings, as shown by the overlay in Figure 2e. This stacking of Er – COT – K – COT forms uniaxial rows of densely packed K[Er(COT)$_2$] complexes. The areal molecular density is estimated to be 1.6 molecules/nm$^2$. The unit cells of both domains (1 and 2) form a rhombic shape with $a_1$ = 0.74 ± 0.05 nm, $a_2$ = 0.88 ± 0.04 nm and $\theta$ = 75°, as shown in Figure 2e (on domain 2 the unit cell is mirrored with respect to $a_2$). However, the vectors of the unit cells are not aligned along any of the principal crystallographic directions of the substrate. Finally, in some areas, small agglomerates of other molecules are found as shown in Figure 2f. The round-shaped molecules have a diameter of 0.90 ± 0.05 nm as highlighted by a green circle, and the size is consistent with the (18-c-6) crown ether of the source material (see also Methods).

## 2.3. Orientation of self-assembled complexes

Linearly polarized X-rays probe empty orbitals of the absorbing atom oriented along the oscillating electric field of the X-rays. Thus, a symmetry breaking of the system gives rise to a difference between absorption spectra measured with different linear X-ray polarizations, the so-called X-ray linear dichroism (XLD). Similarly, circularly polarized X-rays can probe in an element-specific way the projection of the sample magnetization onto the beam direction. By taking the difference of two absorption spectra with opposite circular polarization the X-ray magnetic circular dichroism (XMCD) is obtained.[51]

The XLD and XMCD measurements are complemented by simulations based on a point-charge model implemented in the multiX software.[39] The point charges were put at the $D_{8h}$-symmetrized positions of the carbon atoms as explained in the SI. In particular, the detailed shape of the triple-featured Er$^{3+}$ M$_5$-edge, which depends on the orientation of the X-ray polarization and the incidence angle with respect to the main axis of the SIMs, is reproduced in the calculations.

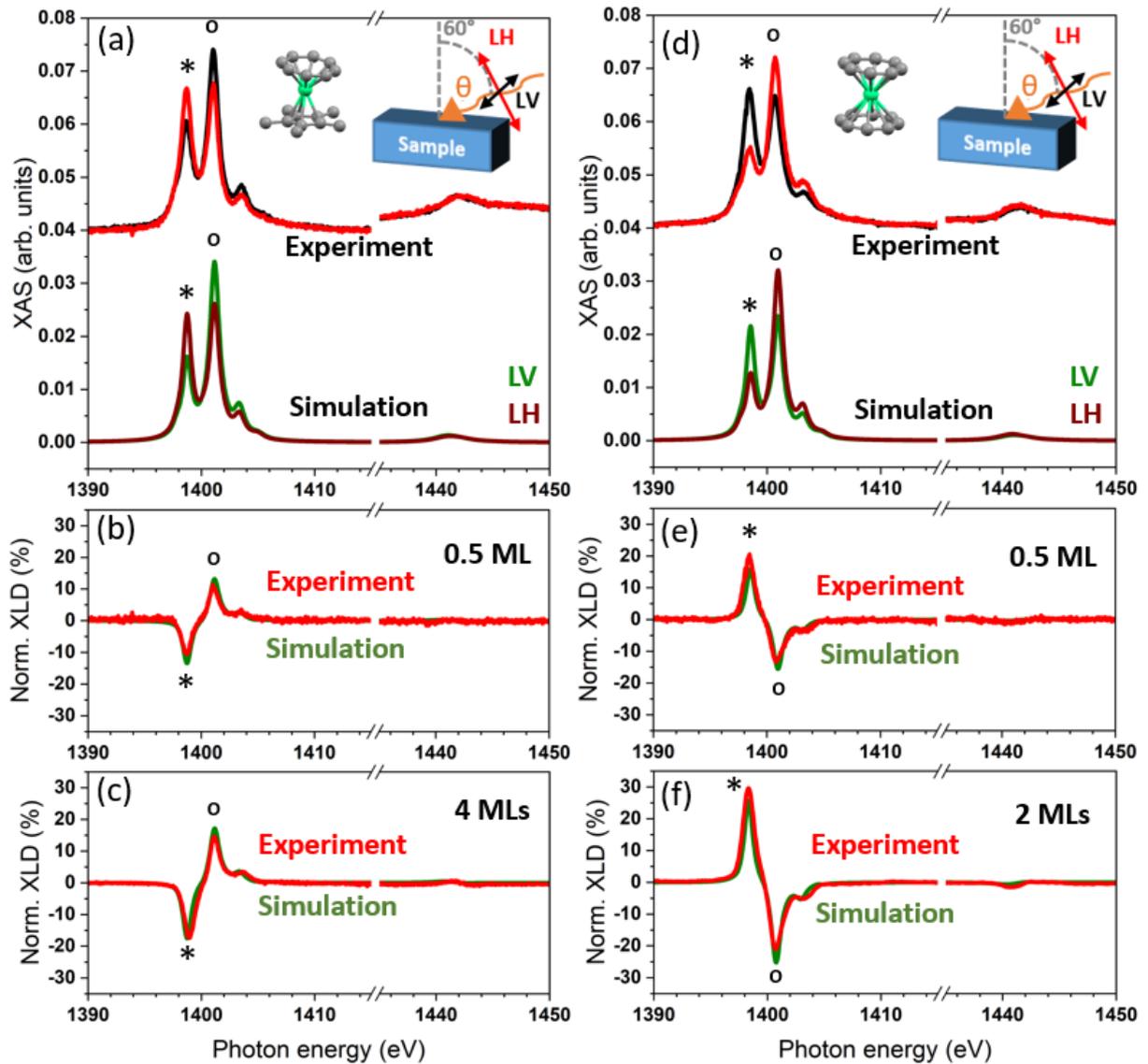

**Figure 3.** Linearly polarized XAS recorded at the Er $M_{4,5}$-edges at 3 K and 50 mT compared to multiX-simulated spectra of (a) Cp*ErCOT(0.5 ML)/Ag(100) and (d) K[Er(COT)$_2$](0.5 ML)/Ag(100). The insets show the molecular structures (H atoms omitted) and the experimental geometry of the grazing-angle measurements, along with the polarization and the beam direction. Panels (b), (c) and (e), (f) show the experimental and simulated XLD spectra of different samples as indicated in the plots. The symbols * and ° identify the main features as described in the main text.

In **Figure 3**, linearly polarized XAS and XLD spectra are reported. Cp*ErCOT(0.5 ML)/Ag(100) (see Figure 3a) shows a similar absorption intensity for the two peaks identified as * (1398.8 eV) and ° (1401.1 eV) in the out-of-plane direction of the substrate (LH, red curve), while the in-plane polarized absorption (LV, black curve) is more intense in the ° peak. The simulation shows that the intensities of the * and ° peaks are given by the orientation of the complex with respect to the incoming photons and their polarization (see SI). Based on the results of the STM images, we have reproduced the spectra by using a model with an equal amount of complexes in the standing-up and lying-down configurations. In the case of the lying-down configuration, the main molecular axis was taken parallel to the substrate plane with uniform disorder over all azimuthal angles. Manual adjustment of the point-charge values of the C atoms ($q_C$) yielded the best agreement between simulated and experimental spectra when fixing the value to $q_C = 0.25\ e$. Figure 3b shows the comparison of the experimental and simulated XLD, with the intensities reported in **Table 1**. The good agreement of the XLD shape points to the fact that the negative sign of * and the positive sign of ° are characteristic of a mixed standing-up and lying-down configuration of the complexes on the surface. The excellent match of the simulated XLD with the experimental ones suggests that both adsorption configurations occur with a ratio of 1:1.

Multilayer samples show a similar shape of the XLD as shown in Figure 3c. While the XAS of the 4 ML sample reported in the SI (Fig. S6) is very similar compared to the sub-ML sample, the spectra are better reproduced by assuming that 55% of the complex are in the standing-up configuration, in good agreement with the experimental results presented in Figure 3 and Table 1.

**Table 1.** XLD intensities of * (~1398.5 eV) and ° (~1401 eV) peaks of experimental and simulated spectra (multiX) of Cp*ErCOT/Ag(100) and K[Er(COT)$_2$]/Ag(100), as reported in Figure 3 and S7.

| Cp*ErCOT | XLD * (%) | | XLD ° (%) | |
|---|---|---|---|---|
| | Experiment | multiX | Experiment | multiX |
| 0.5 ML | -11 ± 2 | -13.4 | 11 ± 2 | 13.1 |
| 4 MLs | -17 ± 3.5 | -17.5 | 15 ± 3 | 17.2 |
| K[Er(COT)$_2$] | | | | |
| 0.5 ML | 21 ± 4 | 15.8 | -13 ± 3 | -15.5 |
| 2 MLs | 30 ± 6 | 25.8 | -21 ± 4 | -25.2 |

On the contrary, K[Er(COT)$_2$](0.5 ML)/Ag(100) shows a similar absorption intensity for both * (1398.3 eV) and ° (1400.6 eV) peaks in the substrate plane (black curve), as reported in Figure 3d. There is a strong absorption asymmetry (red curve) in the out-of-plane direction, with the main contribution coming from the ° peak. This result agrees with the adsorption almost purely in the lying-down configuration, as revealed by the simulations, which reproduce very well the experimental XAS and XLD reported in Figures 3d and 3e. Note that a good fit of the XLD in Figure 3e and of the XMCD spectra shown further below can only be obtained by taking into account a fraction of 13% of the complexes adsorbed in the standing-up geometry. This can be rationalized by assuming that in the sub-ML coverage not all complexes are engaged in the self-assembled islands. In analogy to other complexes containing similar organic macrocycles the isolated molecules prefer the standing-up geometry with the rings parallel to the metal surface plane.[42] The 2 MLs K[Er(COT)$_2$]/Ag(100) sample exhibits the same shape of the XLD as visible in Figure 3f (see also Fig. S6), with an overall larger intensity (see Table 1). Here, no standing-up complexes were assumed in the simulations.

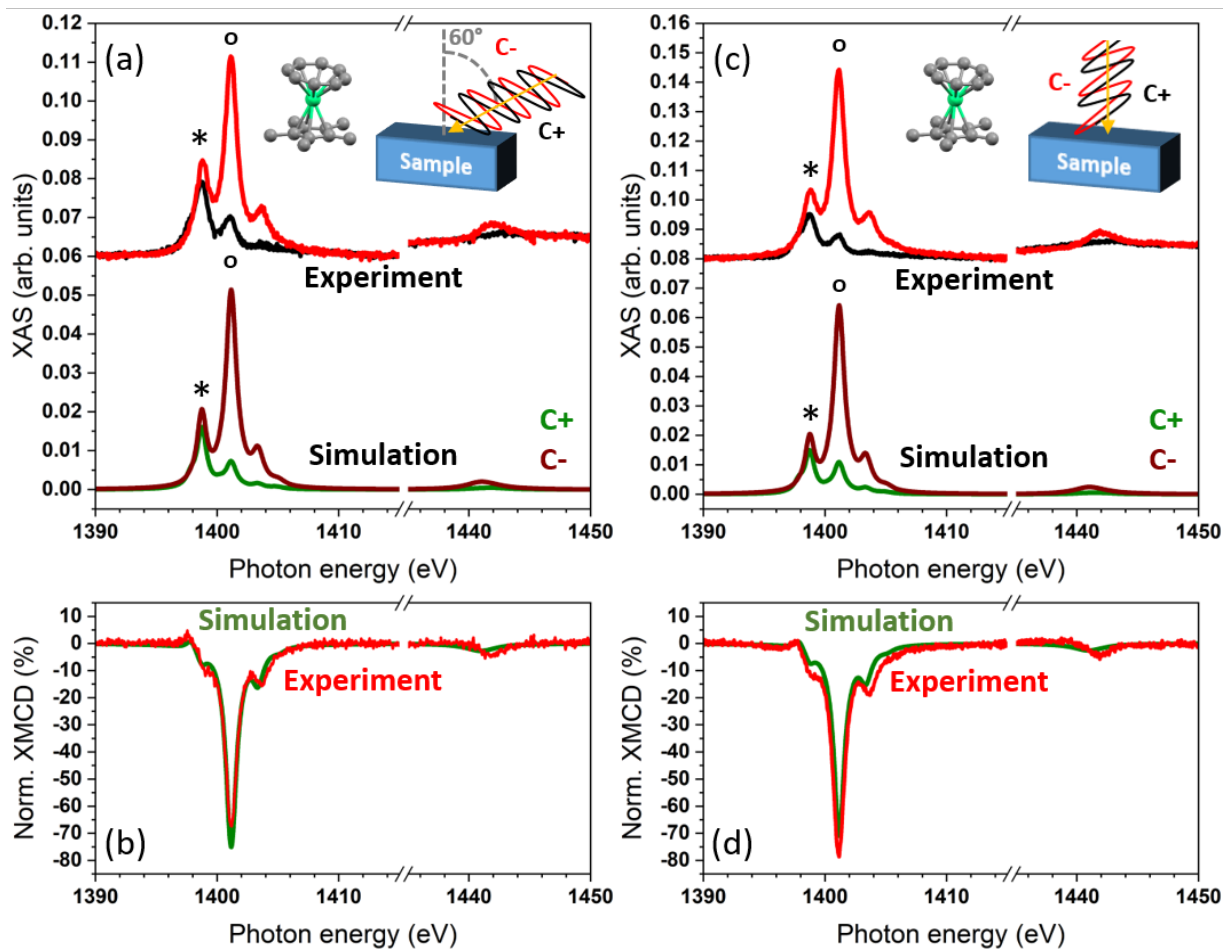

**Figure 4**. Circularly polarized X-ray absorption and XMCD spectra measured at the Er M$_{4,5}$-edges at 3 K and 6.8 T, along with multiX-simulated spectra of Cp*ErCOT(0.5 ML)/Ag(100). (a,b) Grazing (60°) incidence and (c,d) normal incidence as shown in the insets.

The strong easy-axis type magnetic anisotropy of the individual complexes can be used to rationalize the surface ordering by directly comparing XMCD intensities and thus the magnetic moments in normal and grazing incidence. **Figure 4** shows circularly polarized XAS and XMCD of Cp*ErCOT(0.5 MLs)/Ag(100). By using the previously explained multiX model with the equal standing-up versus lying-down ratio, the simulated XAS reproduces well the experimental results in the Figure. The XMCD spectra in Figure 4b and 4d are also well reproduced, with the intensities of the strongest (negative) peaks at 1401.2 eV reported in **Table 2**. For the multilayer, the XMCD spectra of both the normal and grazing incidence become comparable in intensity. The simulations based on the 55:45 standing-up vs. lying-down molecular arrangement described in the XLD section perfectly fit the results in Table 2.

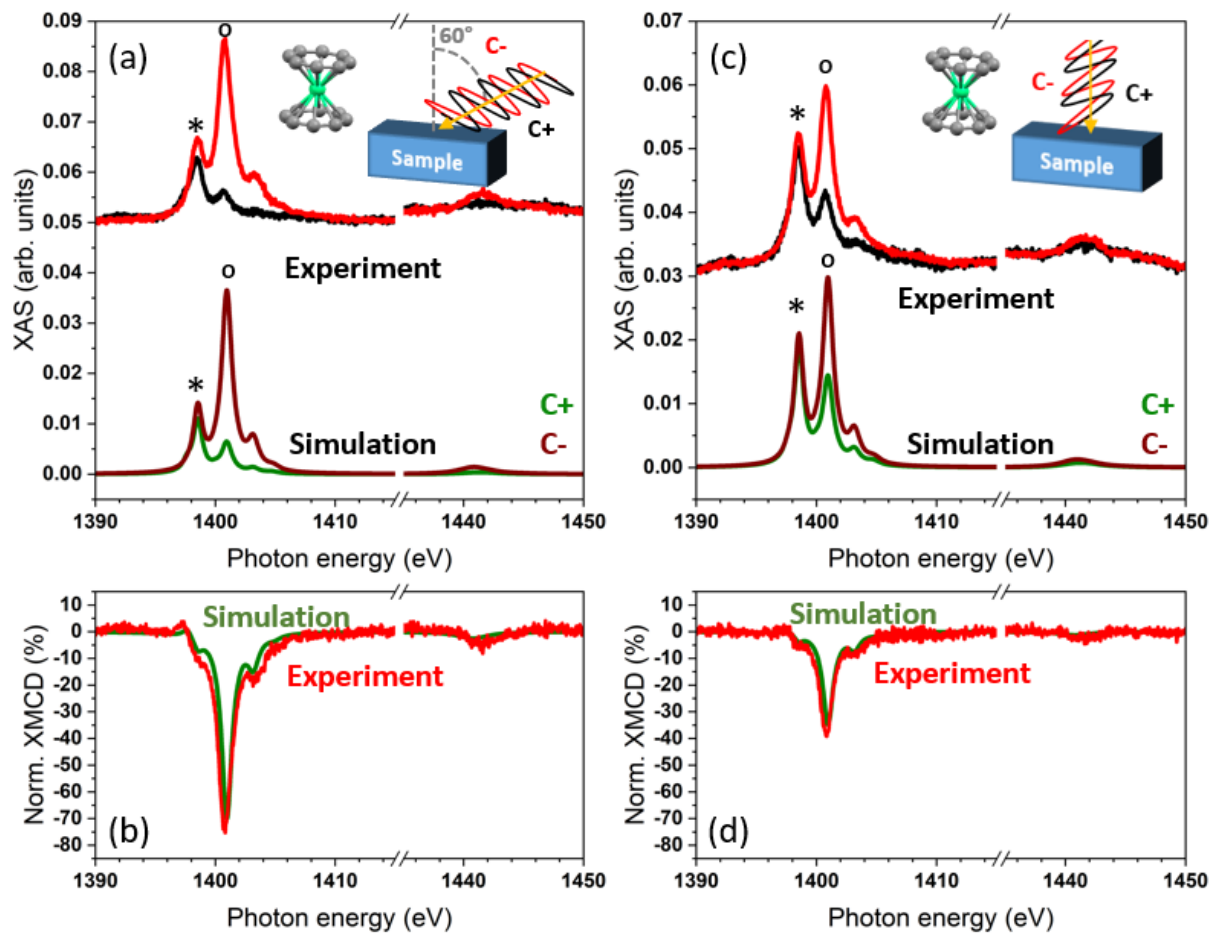

**Figure 5**. Circularly polarized XAS and XMCD recorded at the Er $M_{4,5}$-edges at 3 K and 6.8 T, along with multiX-simulated spectra of K[Er(COT)$_2$](0.5 ML)/Ag(100). (a,b) Grazing (60°) and (c,d) normal incidence as shown in the insets.

Opposite to the Cp*ErCOT SMMs, the circularly polarized spectra of K[Er(COT)$_2$](0.5 ML)/Ag(100) show a strong difference between grazing and normal incidence of the X-rays, as presented in **Figure 5**a and 5c. While the ° peak of the C$^-$ polarization is the dominant one in both cases, it is far less intense in normal incidence. As introduced previously, the best-fit simulations are based on the model with 87% of the complexes self-assembled in the lying-down configuration, isotropically oriented over the azimuthal angles, and 13% in the standing-up geometry. There is excellent agreement between the simulations and the experimental spectra as manifested in the spectra shown in Figure 5, as well as in the peak heights collected in Table 2.

Increasing the thickness of the sample to 2 MLs does not affect much the shape of the XAS compared to the sub-ML (see Fig. S7 of the SI), but the XMCD intensity in normal incidence drastically decreases. This is ascribed to the absence of standing-up complexes in the multilayer sample.

**Table 2**. Experimental and simulated XMCD intensities of Cp*ErCOT and K[Er(COT)$_2$] spectra reported in Figure 4, 5, S7 and S8.

| Cp*ErCOT | XMCD normal (%) | | XMCD grazing (%) | |
|---|---|---|---|---|
| | Experiment | multiX | Experiment | multiX |
| 0.5 ML | 87 ± 9 | 71.0 | 65 ± 7 | 75.0 |
| 4 MLs | 74 ± 7 | 74.6 | 75 ± 8 | 75.6 |
| K[Er(COT)$_2$] | | | | |
| 0.5 ML | 39 ± 4 | 34.7 | 75 ± 8 | 69.9 |
| 2 MLs | 16 ± 2 | 16.4 | 73 ± 7 | 68.0 |

The sum rule analysis[52,53] of the XMCD spectra was performed to obtain the net spin and orbital magnetic moments of the erbium ion. The results are summarized in Table S2 and discussed in the SI. The Cp*ErCOT(0.5 ML)/Ag(100) shows a total magnetic moment of 5 ± 1 $\mu_B$ in normal incidence, about half of the expected value for the pristine complex, attributed to the mixed ordering and azimuthal disorder of the complex. For K[Er(COT)$_2$] the total magnetic moment of 4.5 ± 0.8 $\mu_B$ in grazing against the 1.9 ± 0.8 $\mu_B$ in normal incidence fits the easy-plane type anisotropy of the net magnetization resulting from the in-plane, lying-down, ordering. The XAS and XMCD obtained on the powder sample of the starting material [K(18-c-6)][Er(COT)$_2$]·2THF (*cf.* Fig. S9) exhibit the same shapes as observed on the surface-adsorbed complexes. Furthermore, the corresponding sum rule analysis yields a total magnetic moment of 4.3 ± 0.5 $\mu_B$ consistent with the randomly oriented anisotropy axes of the complexes in the powder as discussed in the SI.

### 2.4. Magnetic hysteresis loops

Figure 6a displays the magnetic hysteresis of Cp*ErCOT(0.5 ML)/Ag(100), measured where only a small hysteresis opening is visible for both incidence angles. The hysteresis loops of the 4 ML samples (Figure S9) are of the same shape as the ones of the sub-ML sample. However, the saturation value is different (see Table S2). Since bulk Cp*ErCOT shows a substantial butterfly-like hysteresis opening at 3 K,[54] the reduction of the hysteresis opening of the surface adsorbed Cp*ErCOT can be attributed to the interaction with the substrate, as discussed below.

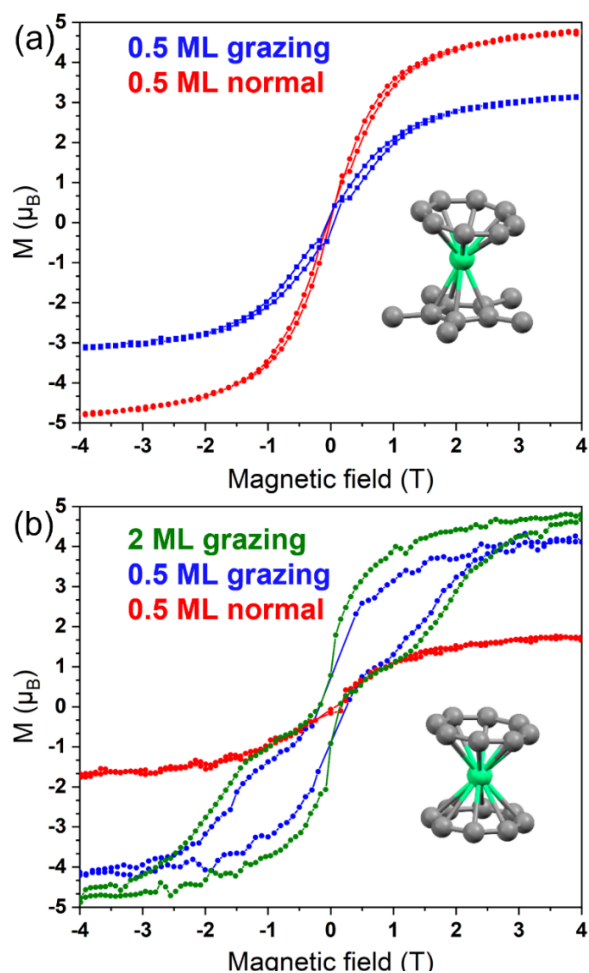

**Figure 6**. XMCD-detected magnetic hysteresis loops recorded at 3 K of (a) K[Er(COT)$_2$] and (b) Cp*ErCOT adsorbed on Ag(100), at a rate of 2 T/min. Normal stands for out-of-plane direction, while grazing (60°) is mostly in-plane. Magnetization values are extracted by sum rule analysis. Lines connecting experimental points are guides to the eyes.

The K[Er(COT)$_2$](0.5 ML)/Ag(100) sample in **Figure 6**b shows a large butterfly-like hysteresis opening between ±3.5 T in grazing incidence, while the loop is essentially closed in normal incidence. This is in agreement with the ordering of the net easy-axis of the magnetization of the complexes parallel to the substrate plane. The different values of the magnetic moments in the two directions are another clear evidence of the net magnetic anisotropy of the sample due to the surface ordering of the complexes, as discussed in the previous section. The hysteresis of the multilayer sample exhibits a very similar, but larger opening in grazing incidence of the X-rays (see also Figure S11). For both coverages, the coercive field was estimated to be $B_{coerc}$ = 0.15 ± 0.07 T, while the remnant magnetization as $M_{rem}$ = 0.84 ± 0.06 μ$_B$. The fact that the hysteresis loop of the 2 MLs sample shows a larger loop opening indicates that the multilayer sample has greater magnetic stability.

## 2.5 Discussion

Cp*ErCOT and K[Er(COT)$_2$] in the sub-ML to ML range exhibit completely different self-assemblies on Ag(100). Cp*ErCOT forms domains of alternating rows of standing-up and lying-down complexes as shown by STM and XAS. A similar ordering was observed for metallocenes on metal substrates, where such a configuration is dictated by the T-shaped van der Waals interaction between the Cp ligands of neighboring molecules.[49,55] The presence of lying-down nickelocene was shown to be necessary to minimize the adsorption energy of the molecules on metals,[55] and it is likely the case also for the Cp*ErCOT studied here. The main difference compared to the metallocenes is that the lying-down Cp*ErCOT complex connects "diagonally" two standing-up complexes of neighboring rows rather than perpendicularly to the standing-up rows of complexes, reproducing the ordering formed along the [-101] direction in the Cp*ErCOT unit cell. Also, the orientation of lying-down rows changes between consecutive rows, forming a herringbone-like pattern, as shown in Figure 2a and 2c. The herringbone structure is found also in the compact configuration of metallocenes,[49] caused by the lateral shift of 0.15 nm by standing-up complexes belonging to every second row, which is also observed in Cp*ErCOT/Ag(100). This likely happens so the complexes can accommodate in the energetically favorable adsorption sites on the surface, while still being influenced by the intermolecular interaction.

On the other hand, heteroleptic sandwich complexes can deviate from this ordering, as recently reported for CpTiCOT/Au(111).[44] In that case, the compound forms a complex unit cell of mixed standing-up and lying-down complexes in a 1:3 ratio, due to stronger intermolecular interactions. In the present case STM, XLD and multiX simulations suggest that equal amounts of the two Cp*ErCOT conformations are present. Although the model used is very simple and based on the highly symmetric D$_{8h}$ symmetry of the point charges for both compounds, it can reproduce the experimental XLD and XMCD spectra with great fidelity. Most importantly, it points out that the reversed sign of the XLD shapes of the two complexes is given by their different orientations. In fact, K[Er(COT)$_2$] forms domains of highly oriented rows with lying-down complexes and with the main molecular axis oriented along the same direction. Such a conformation suggests that the [Er(COT)$_2$]$^-$ anions are conjugated through the intercalation of K$^+$ ions, in a similar manner as reported for the K(THF)$_4$[Er$_2$COT$_4$] tetralayer.[34] Indeed, the STM-measured distance between two K[Er(COT)$_2$] complexes is $a_2$ = 0.88 ± 0.04 nm (*cf.* Figure 2f), which coincides with the Er-Er distance in the K(THF)$_4$[Er$_2$COT$_4$] cluster. Altogether, this suggests that K[Er(COT)$_2$] forms rows of alternating Er$^{3+}$ and K$^+$ ions spaced by standing COT$^{2-}$ rings. A similar linear wire-like configuration of COT-based single-chain magnets on surfaces was reported for EuCOT/Graphene/Ir.[50,56] For neutral π-conjugated systems, a preferential ordering with the rings parallel to the surface plane or driven by the T-shaped interaction would be expected as discussed previously. In contrast, in the case of K[Er(COT)$_2$] the charge balance *via* intercalation of a potassium ion between two complexes is the main mechanism driving the self-assembly.

The vertical position of the K$^+$ ions in the first layer is likely influenced by their charged nature: They adsorb closer to the metal surface, instead of being centered between two (COT)$^{2-}$ rings, as in the reported tetralayer.[34] This creates a vertical charge separation with a positive layer on the surface, balanced by the metal surface, and a negatively charged layer of [Er(COT)$_2$]$^-$ anions. The charge separation in the sub-monolayer case forms an electric dipole that acts as a built-in potential, shifting the positions of the XPS peaks, as reported for other dipolar materials.[57,58] By increasing the thickness of the sample, the influence of the surface on the potassium cations becomes less important and more axially aligned complexes are formed. This effect explains the rather low C 1s binding energy of 283.9 eV of the sub-ML K[Er(COT)$_2$] and the core level shifts reported in the SI, although the contribution of surface screening cannot be excluded. A depiction of the adsorption conformations of the studied complexes is shown in Figure S3.

We conclude that Cp*ErCOT experiences a stronger interaction with the surface based on three observations. 1) The presence of two different sp$^2$ C 1s signals at binding energies of 284.9 eV and 285.9 eV, representing the non-interacting carbon atoms of the aromatic rings and the ones interacting with the metal substrate (*cf.* Fig. 1): Considering a standing-up and a lying-down complex, out of 36 carbon atoms 8 are hybridized with the metal surface, because the COT ring of the standing-up complex has its π-electron cloud in the metal plane; 18 sp$^2$ carbon atoms, the inner ones of the Cp* ligand and the remaining COT of the lying-down species, contribute to the peak at lower binding energy; the 10 sp$^3$ carbons, which are part of the methyl groups of the Cp* ligands give rise to an intermediate peak at 285.2 eV, typical for sp$^3$ carbon.[45] Consistently, the C 1s peak areas shown in Figure 1c exhibit a 8:10:18 ratio. This scenario is very similar to the interpretation in ref. [45], in which two different C 1s signals at similar binding energies as in the present text were observed (with the exclusion of the sp$^3$ peak). 2) The substrate templating effect: The unit cell vectors of the molecular Cp*ErCOT overlayer coincide with the main crystallographic directions of the substrate. 3) The shrinkage of the hysteresis loops of the surface-adsorbed complexes compared to the bulk phase.[29] Our measurements show that the surface-supported complexes display a faster magnetization relaxation, considering the faster sweeping rate of the magnetic field used here. The stronger interaction of the Cp*ErCOT complexes results from the hybridization of the π orbitals of the standing-up complexes directly with the metal, which breaks the symmetry of the ligand-field and makes the complexes prone to vibronic coupling and dynamic charge transfer with the metal, all of which enhance the magnetization relaxation.[19–24] The X-ray-induced demagnetization has been estimated to give a minor contribution to the closing of the magnetic hysteresis, given the low photon flux and the molecular density, as compared to previous studies and K[Er(COT)$_2$].[26,59]

Contrarily, K[Er(COT)$_2$] exhibits a strong net magnetic anisotropy due to the orientation of the molecular easy axes parallel to the substrate plane. The hysteresis opening proves clearly the slow relaxation of the magnetization, resembling the butterfly opening reported for the bulk phase of [K(18-c-6)][Er(COT)$_2$]·2THF.[33,38] A direct comparison of the hysteresis loops is challenging because of different field sweep rates. However, following up on the literature on the bulk phase we ascribe the temperature-assisted quantum tunneling of magnetization process to be the dominant one, due to the similarities in the low field region. Yet, also here X-ray induced demagnetization is expected to give a small contribution at low magnetic fields.

In the present study, the different hysteretic behavior of the two complexes is attributed to the orientation of the ligands' π orbitals with respect to the substrate which allows or suppresses hybridization and thus favors or suppresses vibrational coupling to the Ag(100) substrate. The direct interaction of the π orbitals of standing-up Cp*ErCOT with the substrate induces faster magnetization relaxation. On the other hand, because of the preferentially lying-down orientation of K[Er(COT)$_2$] the π orbitals are parallel to the surface plane and thus weakly interacting with the substrate. Importantly, due to this mechanism, the orientation of the complexes has an enormous impact on the magnetic relaxation properties.

## 3. Conclusions

The present study shows how structurally similar lanthanide organometallic SIMs based on π-conjugated ring-shaped ligands can exhibit completely different self-assemblies on metal surfaces, due to different intermolecular interactions, as well as different degrees of interaction with the Ag(100) substrate. Cp*ErCOT forms compact rows of alternating standing-up and lying-down complexes along the [010] and [001] directions of the substrate. K[Er(COT)$_2$] arranges in highly ordered domains with complexes purely in the lying-down conformation, stabilized by the coordinated K$^+$ ion. Polarized XAS indicates that the net magnetic anisotropy is strong for surface-adsorbed K[Er(COT)$_2$] and weak for Cp*ErCOT. Also, Cp*ErCOT exhibits weak hysteresis opening both in-plane and out-of-plane because of the stronger interaction of the standing-up complexes with the substrate through their conjugated π orbitals. Contrarily, the hysteresis opening of K[Er(COT)$_2$] is large, suggesting a minor influence of the substrate on the magnetic relaxation properties. This is linked to the weak interaction of the π-conjugated ligand orbitals with the substrate, given their orientation parallel to the substrate plane. This suggests that the π electron cloud directly interacting with the metallic surface results in an increased magnetization relaxation rate, possibly through enhanced vibronic molecule-substrate coupling, opposite to the case with the π orbitals oriented parallel to the metal substrate. This result is of great importance for the design of future surface-adsorbed molecular architectures aiming to achieve stable and addressable molecular magnets.

## 4. Experimental Section

*Synthesis of SIMs*: the Cp*ErCOT and K(18-c-6)Er(COT)$_2$·2THF complexes were synthesized as described in the literature.[29,33]

*Sample preparation*: The air-sensitive complexes were handled in He gas environment of a glovebox (H$_2$O/O$_2$ < 1 ppm). The samples were obtained in the range between a sub-monolayer and a few monolayers by thermal sublimation of bulk crystallites on freshly prepared single-crystal Ag(100) surface in UHV. Polycrystalline powder of [K(18-c-6)][Er(COT)$_2$]·2THF was used as a source to deposit the K[Er(COT)$_2$] complex, after degassing the (18-crown-6) and THF molecules below the sublimation temperature of 360 °C used for the samples shown in the text. The molecular coverage of the samples was estimated by a quartz crystal microbalance, calibrated by STM. The extracted coverage was cross-referenced with the adlayer thickness obtained from the attenuation of the Ag 3d levels of the substrate measured by XPS. Further details are reported in the SI.

*XPS:* Survey scans were performed by using monochromatic (Δ$E$ = 0.1 eV) synchrotron light at 800 eV at the PEARL beamline at the SLS, Paul Scherrer Institut.[60] A front-end aperture of 3 × 4 mm$^2$, an exit slit of 30 μm and a dwell time of 0.5 s per step were used. Detailed scans were performed with energy steps between 0.03 eV and 0.5 eV.

*STM*: Images were acquired at 4.5 K using an Omicron LT-STM and post-processed with Gwyddion.[61] Atomically resolved scans of the bare Ag(100) surface were acquired for reference.

*XAS*: The XAS, XLD and XMCD spectra were acquired in total electron yield mode using on-the-fly scans at the EPFL/PSI X-Treme beamline at the SLS.[62] To avoid beam-induced damage of the samples a defocused beam and a reduced photon flux were employed by minimizing the exit slit opening. The average photon flux used to measure the *M(H)* curves of the Cp*ErCOT samples was $2 \times 10^{-2}$ ph nm$^{-2}$ s$^{-1}$, while the maximum flux used for K[Er(COT)$_2$] was $5 \times 10^{-2}$ ph nm$^{-2}$ s$^{-1}$. The applied magnetic field was always collinear with the X-ray beam. To obtain hysteresis loops, the two circular polarizations were recorded separately for the two sweep directions. The intensity difference between the absorption at the energy of maximum XMCD and the pre-edge was evaluated while sweeping the magnetic field at a rate of 2 T min$^{-1}$. The *M(H)* curves were rescaled to the total magnetic moments extracted by the sum rule analysis. XAS is defined as the sum of the two polarized absorption spectra ($\mu_V + \mu_H$) or ($\mu^+ + \mu^-$). The XLD is defined as the difference $\mu_V - \mu_H$, while the XMCD as $\mu^+ - \mu^-$.

*XAS Simulations*: Details of the simulations performed using the multiX software are reported in the SI.

**Supporting Information**

Supporting Information is available from the Wiley Online Library or from the authors.

**Acknowledgments**

V.R. and J.D. gratefully acknowledge funding from the Swiss National Science Foundation (grant no. 200020_182599). M.B. would like to thank the ETH Zürich Grant program (ETH-44 18-1). N.D. acknowledges funding by the Swiss National Science Foundation (grant no. PZ00P2_193293). M.M. thanks the University of Ottawa, the Canada Foundation for Innovation and the Natural Sciences and Engineering Research Council for financial support. M.H. acknowledges funding from the Swiss Nanoscience Institute.

Author Contributions

**V.R.** (data curation: lead; investigation: lead; validation: lead; visualization: lead; writing‐original draft: lead); **M.B.** (resources: equal; writing‐review & editing: supporting); **M.H.**

(investigation: supporting; writing‐review & editing: supporting); **D.V.** (investigation: supporting; writing‐review & editing: supporting); **K.H.** (resources: supporting; writing‐review & editing: supporting); **N.D.** (investigation: supporting; writing‐review & editing: supporting); **B.D.** (investigation: supporting; software: lead; writing‐review & editing: supporting); **M.D.K.** (resources: supporting; writing‐review & editing: supporting); **M.Muntwiler** (resources: supporting; supervision: supporting; writing‐review & editing: supporting); **C.C.** (funding acquisition: supporting; resources: supporting; supervision: supporting; writing‐review & editing: supporting); **M.Murugesu** (funding acquisition: supporting; resources: supporting; Supervision: supporting; writing‐review & editing: supporting); **F.N.** (supervision: supporting; writing‐review & editing: supporting); **J.D.** (funding acquisition: lead; project administration: lead; supervision: lead; writing‐review & editing: lead).

**Data Availability Statement**

The presented data are available from the authors on request.

**Conflict of Interest**

The authors declare no conflict of interest.

# Supporting Information

## Orientation-Driven Large Magnetic Hysteresis of Er(III) Cyclooctatetraenide-Based Single-Ion Magnets Adsorbed on Ag(100)

*Vladyslav Romankov, Moritz Bernhardt, Martin Heinrich, Diana Vaclavkova, Katie Harriman, Niéli Daffé, Bernard Delley, Maciej Damian Korzyński, Matthias Muntwiler, Christophe Copéret, Muralee Murugesu, Frithjof Nolting, Jan Dreiser\**

## Table of Contents

# 1. Sample preparation

Polycrystalline powders of [K(18-c-6)][Er(COT)$_2$]·2THF and Cp*ErCOT were used as starting materials.[1,2] Quartz crucibles were filled with ~10 mg of powder and transferred to the preparation systems of the X-Treme[3] and PEARL[4] beamlines. Due to the high oxygen and moisture sensitivity, the materials were handled in the inert He environment of a glovebox. The degassing and sublimation were performed by using a commercial organic effusion cell (Kentax) at the X-Treme beamline and a custom-made multi-pocket evaporator at PEARL. The complexes were deposited on the (100) surface of an Ag single crystal freshly prepared by Ar$^+$ sputter-annealing cycles. The deposition rate was measured using a quartz crystal microbalance.

# 2. Stoichiometric characterization of K[Er(COT)$_2$]

The yellow-colored [K(18-c-6)][Er(COT)$_2$]·2THF powder was degassed up to 107°C, after which several control samples were prepared at increasingly higher crucible temperatures of up to 360°C. The shifts of the XPS core levels as a function of the sublimation temperature of the compound and the coverage/thickness of the sample adlayer are reported in **Figure S1**. The background removal was performed by subtraction of a Shirley function in the C 1s and K 2p spectra, as well as spectra of Er 4d level of higher coverage samples (7 and 8 MLs), which display a larger signal-to-noise ratio. In the Er 4d spectra of samples with lower coverage and in all O 1s spectra a straight line was subtracted. The main C 1s core level peaks as shown in Figure S1 were normalized to unity for a better comparison of the features and chemical shifts. The intensities of the other core level spectra are normalized by the same factors used to normalize the C 1s core levels.

When the complex is sublimed at a crucible temperature of 107°C (red curves), a strong presence of carbon at 286.3 eV and oxygen at 532.9 eV is detected, while erbium and potassium are essentially negligible. The peaks are attributed to the deposition of large amounts of (18-c-6) crown ether molecules on the surface, due to the typical binding energy of the oxygen-bound carbon in such molecules[5,6] and the absence of the carbon sp$^2$ peak that would appear in the presence of tetrahydrofuran (THF) and/or COT$^{2-}$.[7]

In the range from 200°C to 320°C (purple curves), we found the presence of C, Er and K. Together with a strong O 1s peak, the presence of these elements and two different carbon environments suggests that the sample has the same elements as the polycrystalline starting material of [K(18-c-6)][Er(COT)$_2$]·2THF. Indeed, the latter compound has 16 aromatic carbon atoms and 4 carbon atoms of the 2 THF molecules not bound to oxygen, while there are 12 oxygen-bound carbons in the crown ether and 4 in the 2 THF molecules. The ~1 ML sample produced at this temperature shows a ratio of the two carbon peak areas of 1.36 (obtained from fits using Voigt functions), which is in good agreement with the ratio of 1.25 expected for the polycrystalline starting material.

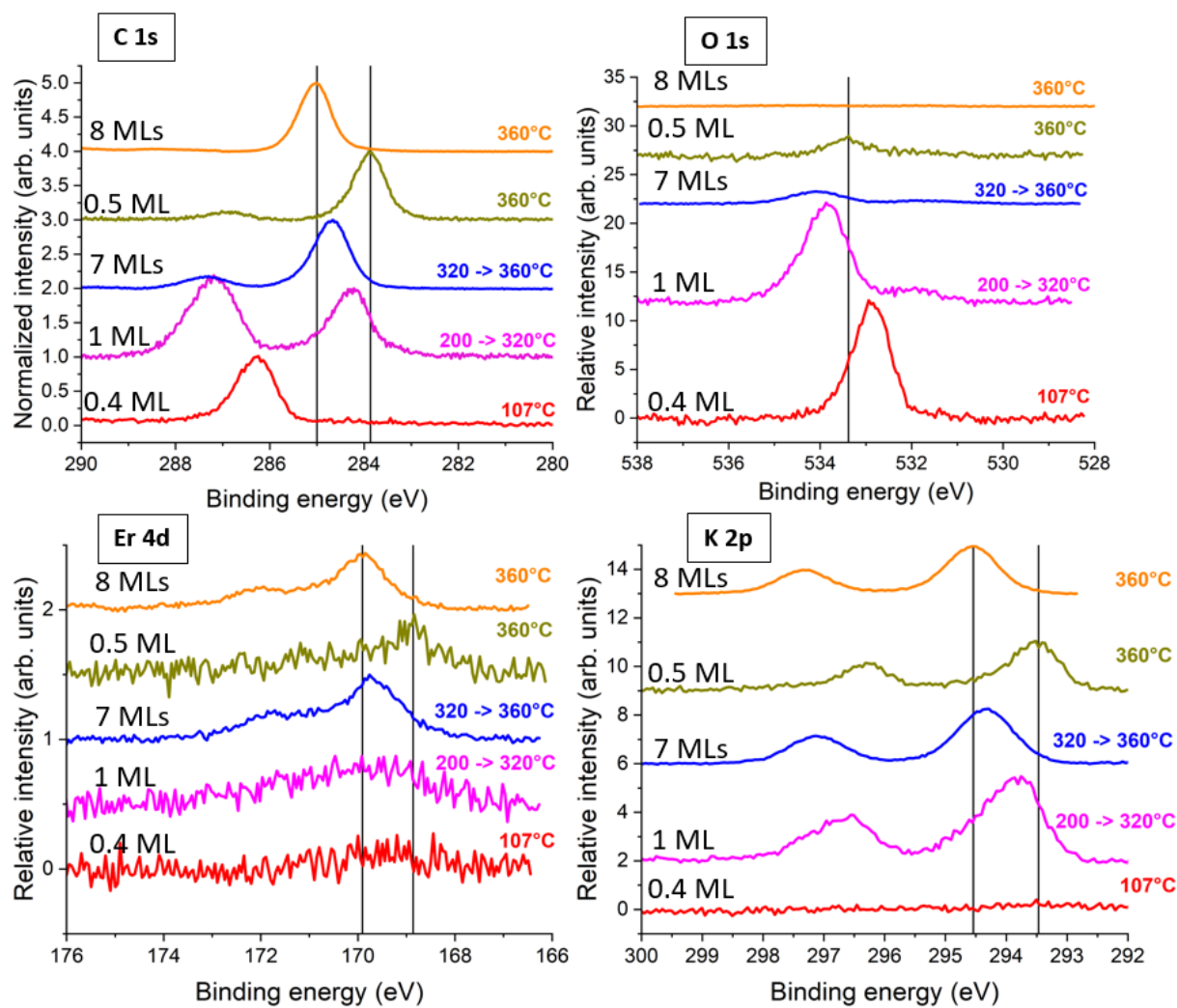

**Figure S1**. X-ray photoelectron spectra of C 1s, O 1s, Er 4d and K 2p core levels of K(18-c-6)[Er(COT)$_2$]·2THF deposited at different crucible temperatures on an Ag(100) substrate. While the C 1s peak is normalized to unity, the other edges are scaled by the same factors used for the C peak. The estimated coverage of the sample is indicated for each spectrum. The black vertical lines indicate the central position of the main core level peaks of the sample reported in the main text (dark yellow).

For crucible temperatures of 320-360°C (blue curves), the dominant type of carbon becomes that of the aromatic rings. The strong Er and K presence, and the weak C-O carbon (at higher binding energy) and oxygen signals suggest that the dominant species on the surface is K[Er(COT)$_2$]. From the blue curve in the C 1s spectra it is possible to deduce that the ratio between the C-O and the sp$^2$ carbon peak areas is only 0.23. Assuming the crown ether to be the only contamination species, this roughly corresponds to a ratio of one crown ether per three K[Er(COT)$_2$] complexes.

At a temperature of 360°C, which has been used to prepare the K[Er(COT)$_2$] samples reported in the main text (dark yellow and orange curves), the presence of potassium, erbium, the (almost) negligible oxygen and mainly the aromatic carbon peak at the C 1s edge suggest that the main species deposited at this temperature is K[Er(COT)$_2$]. This simple analysis is confirmed by the stoichiometry analysis, as reported in the main text. The small C-O carbon signal indicates the presence of impurities on the surface. If all impurities are assumed to be crown ether molecules, the ratio of the C-O and sp$^2$ peaks shows that at most *one* crown ether molecule per *seven* K[Er(COT)$_2$] complexes is present. In the presence of THF or other C and O containing contaminants, the ratio of crown ether vs K[Er(COT)$_2$] would be even lower than 1:7. In the case of the multilayer sample (orange curves) the signals of the contaminants almost vanish. As compared to Er, the area ratios of C, O and K peaks of this sample amount to 20.5, ~0 and 1, which are close to the expected values for K[Er(COT)$_2$] without crown ether. This can be understood by the chronological sequence of the experiments and the subsequent, increasing loss of the crown ether from the crucible as the 0.5 ML sample was prepared before the 8 ML one. The negligible traces of C-O carbon in the C 1s core level spectrum of the 8 ML sample and the absence of the O 1s peak confirm that the (18-c-6) crown ether and the THF molecules are already sublimed at temperatures lower than 360°C.

Moreover, the spectra shown in Figure S1 exhibit a thickness-dependent shift of all core level peaks. While the sub-monolayer samples have the lowest binding energy for all core levels, increasingly thicker molecular layers show a progressive shift of all core level peaks toward higher binding energy. The two boundaries, given by the ~0.5 ML (dark yellow) and the ~8 MLs (orange) coverages, have the core levels shifted on average by 1.05 eV with respect to one another, as indicated by the position of black vertical lines in Figure S1 (excluding oxygen). This can also be seen in **Figure S2**, where the relative shift of all peaks toward lower binding energy vs. estimated coverage is plotted. The shifts have been attributed to the vertical charge separation of the complex at the surface, which causes the formation of a built-in electric potential shifting the core levels toward lower binding energies. We assume that the K$^+$ ions adsorb closer to the substrate because of the charge compensation of the metal surface. Because of the smaller influence of the metal substrate in the multilayer sample the charge separation effect becomes negligible. This concept is depicted in **Figure S3**, where the molecular orientation of the self-assembled complexes on Ag(100) is shown. Indeed, thickness-dependent test samples show a consistent shift of all core levels toward higher binding energies, in line with other polar materials reported in the literature.[8,9] Indeed, the samples with larger coverages tend to recover the position of the core levels, which are more aligned with the data reported for C 1s,[10–12] K 2p[13,14] and Er 4d.[15,16] In particular, the binding energy of the C 1s peak of the thickest test sample (8 MLs) coincides with the carbon peak of the π rings of the Cp*ErCOT(1 ML)/Ag(100) sample. Nevertheless, we do not exclude that the surface-induced screening effect or the charge accumulation of the thicker test samples can contribute to the shifts of the peak positions as well.

On the other side, small shifts of core levels to lower binding energies for sub-monolayer samples can also be attributed to the screening of the core hole performed by the image state electrons of metallic substrates. However, on metal surfaces such shifts are usually not greater than ~0.6 eV,[17] leading to a broadening of the peak, with a tail toward the lower binding energy side. In the case of fullerenes on Cu(111), the molecule-substrate interaction promoted shifts up to ~0.7 eV for the C 1s peak attributed to charger transfer[18], while on Au(100) the reported shift was -0.8 eV[19], similarly to other cases[17,20,21]. However, there is no clear indication of a charge transfer for K[Er(COT)$_2$]/Ag(100), since all the edges shift uniformly toward higher binding energies with increasing coverage.

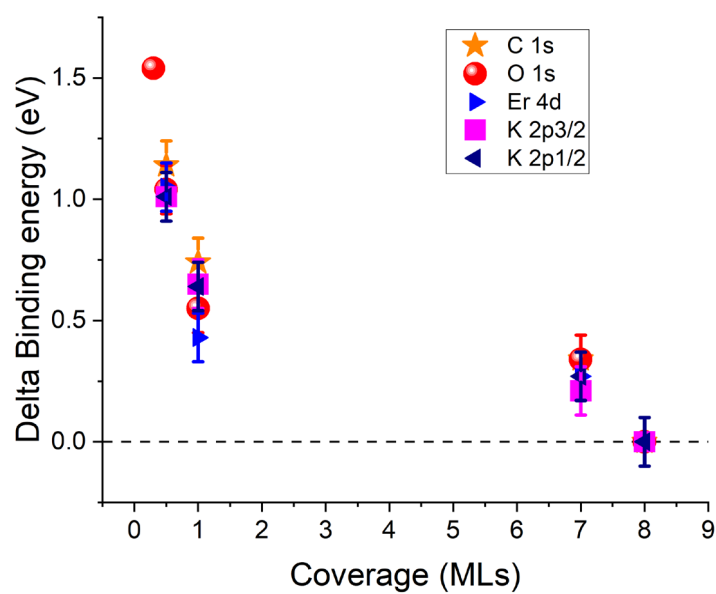

**Figure S2**. Thickness-dependent shift of the XPS core levels of K[Er(COT)$_2$] SMMs deposited on Ag(100), as explained in the text. The error bars are given by the FWHM of the different peaks.

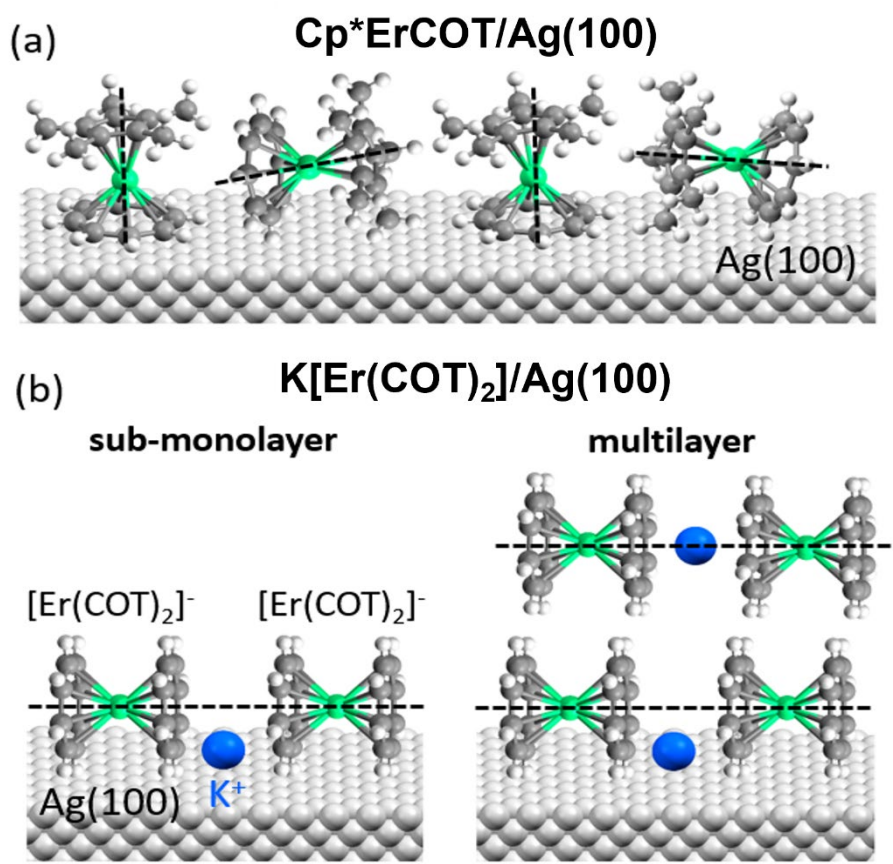

**Figure S3**. Model of (a) Cp*ErCOT/Ag(100) along the [-101] crystallographic direction and (b) K[Er(COT)$_2$]/Ag(100) along the axial direction as explained in the main text. The dashed lines represent the molecular axes of the complexes.

## 3. Additional STM images

In this section, we report two extra STM images of ~1 ML of Cp*ErCOT SMMs deposited on Ag(100). **Figure S4**a shows an area of 100 x 100 nm$^2$ acquired at 50 pA and 0.25 V. Highly oriented rows of molecular complex can be identified by an alternation of brighter and darker stripes in the diagonal direction of the Figure. The self-assembled complex forms multiple domains of rows with the same geometrical configuration, as explained in the main text. At the domain boundaries, the contrast of the rows is inverted, so that brighter rows become darker and vice-versa. Figure S4b shows an area of 500 x 500 nm$^2$ acquired at 50 pA and 0.25 V. Different terraces can be identified in the Figure, due to the step-like structure of the Ag substrate underneath. Line profile scans acquired across the terrace steps give a vertical step of

~0.25 ± 0.05 nm. Figure S4b also shows artifacts due to the mobility/dragging of the complex clusters under the STM tip.

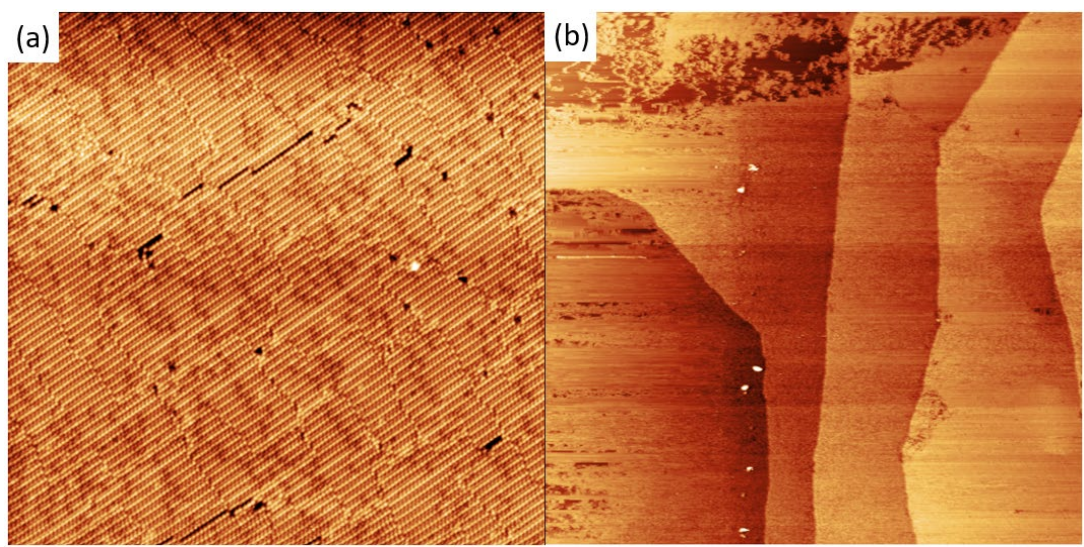

**Figure S4**. Constant-current STM images recorded at 4.5 K of Cp*ErCOT(~1 ML)/Ag(100). The imaging conditions are: (a) 100 x 100 nm$^2$, acquired at 50 pA and 0.25 V; (b) 500 x 500 nm$^2$, 50 pA and 0.25 V.

## 4. MultiX simulations

The multiX software[22] was used to simulate the XAS reported in the main text. In this method, the ligands surrounding the absorbing atom are taken into account by point charges. This simple yet powerful model is useful to understand the main features of the XAS, XLD and XMCD spectra and their dependence on the X-ray incidence angle and the applied magnetic field. An example input file can be found in Section 8 of this Supplementary Information.

The multiX approach relies significantly on first principles to calculate the details of the electronic structure of the central atom and its multiplet structure in a ligand field from very little input. The electron orbitals, which are needed to construct all determinant functions arising from the open shells in the ground and the excited states, are based on a local density approximation (LDA) self-consistent fully relativistic atomic calculation inside multiX. The LDA provides the gross excitation energy. It is only slightly corrected near the 1% level by a semiempirical threshold correction in the preset case. The spin-orbit splitting, setting the separation between M$_{4,5}$ edges, is scaled down by ~5% semiempirically. The electron-electron interaction is scaled down by 15% from the bare orbital-based result. These two scalings are very typical.

The environment of the rare earth atom has a great influence on the details of the atomic ground state. For the present calculations, a simple model with only 2 x 8 carbon atomic positions is used. In principle, the delocalized electron cloud of the π orbitals of the ligand rings and their almost parallel planes can be parameterized as an effective ligand field of $C_{\infty v}$ symmetry, acting on the central $Er^{3+}$ ion.[23] For this reason, in our model we assume that the effective point charges perceived by the $Er^{3+}$ ions of the two complexes are very similar due to the sandwich structure of the systems and, although the coordination environment of the two compounds is different, the effective charge can be parameterized similarly. Hence the simulations are based on a structure with an enforced $D_{8h}$ symmetry, which is realized by positioning accordingly the carbon atoms of the first coordination shell with the average Er-C bond distance taken from literature.[1,24] A single value of carbon point charge as a fit parameter for the crystal field/ ligand field is used as an actual fit parameter. The precise value of this fit parameter has a significant influence on the magnetic properties and is crucially defined by the XLD and XMCD measurements. The nominal positions are reported in **Table S1**. To keep the model simple and the number of parameters low, we manually changed the point charge parameter of the carbon atoms and the fraction of complexes oriented with the molecular axis in-plane or out-of-plane as compared to the substrate plane for both compounds. Furthermore, to match the experimental spectra, the Coulomb and spin-orbit interactions were scaled to 85% and 95% of the computed values, respectively. A core-hole lifetime broadening of 0.45 eV to 1.45 eV in the span of 1398 eV to 1440 eV was used to simulate the peak widths at the $M_{4,5}$-edges. The ligand field was scaled by a multiplication factor of 1.182.

**Table S1**. Atomic positions and charges used to generate the ligand field of a standing-up generic molecule in the $D_{8h}$ symmetry implemented in the multiX code. The positions were extracted from ref. [1].

| Atom | x (Å) | y (Å) | z (Å) | $q_C$ (e) |
|---|---|---|---|---|
| C | 0 | 1.831 | 1.912 | 0.25 |
| C | 1.2947 | 1.2947 | 1.912 | 0.25 |
| C | 1.831 | 0 | 1.912 | 0.25 |
| C | 1.2947 | -1.2947 | 1.912 | 0.25 |
| C | 0 | -1.831 | 1.912 | 0.25 |
| C | -1.2947 | -1.2947 | 1.912 | 0.25 |
| C | -1.831 | 0 | 1.912 | 0.25 |
| C | -1.2947 | 1.2947 | 1.912 | 0.25 |
| C | 0 | 1.831 | -1.912 | 0.25 |
| C | 1.2947 | 1.2947 | -1.912 | 0.25 |
| C | 1.831 | 0 | -1.912 | 0.25 |
| C | 1.2947 | -1.2947 | -1.912 | 0.25 |
| C | 0 | -1.831 | -1.912 | 0.25 |
| C | -1.2947 | -1.2947 | -1.912 | 0.25 |
| C | -1.831 | 0 | -1.912 | 0.25 |
| C | -1.2947 | 1.2947 | -1.912 | 0.25 |

All simulated spectra are based on linear combinations of three configurations of the model complex giving rise to different absorption spectra: the standing-up configuration, with the axis normal to the surface plane and the two lying-down configurations, with the axis rotated by 90° around the x- and y-axis. Since the grazing incidence was imposed by a tilt angle of 60° with respect to the y-axis, the in-plane molecules were simulated by the average of the spectra of the two lying-down configurations, oriented with the axis of the complex parallel to the x and y reference axes. The experimental conditions, i.e., the magnetic field and temperature, were considered: The magnetic field was simulated parallel to the X-ray beam direction. Its value was 50 mT and 6.8 T for the linearly and circularly polarized XAS, respectively.

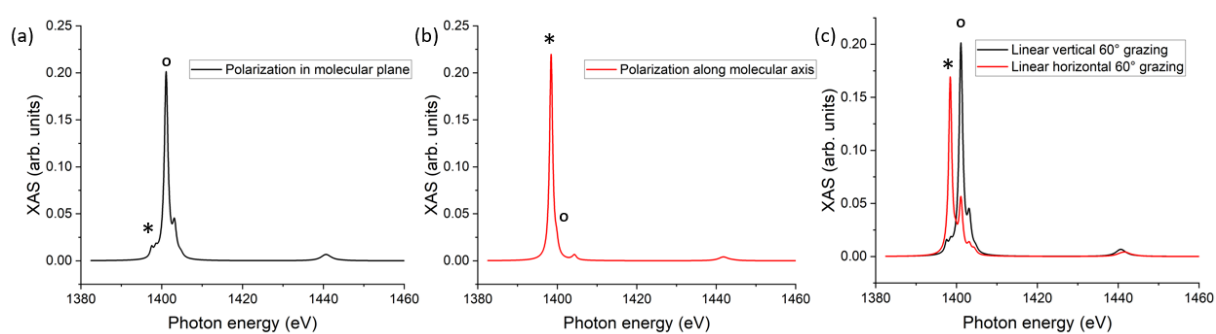

**Figure S5**. MultiX-simulated XAS at the Er $M_{4,5}$-edges as explained in the text. The spectra are shown for the molecular axis (a) perpendicular and (b) parallel to the molecular axis of the model system. (c) A spectrum obtained by using two perpendicular polarization vectors oriented at a grazing angle of 60° with respect to the molecular axis.

A model calculation for the case of linearly polarized X-rays is reported in **Figure S5**. When the polarization vector is oriented in the plane of the simulated $COT^{2-}$ ligands, the absorption spectra show a strong peak around 1401 eV ("o" feature) and negligible intensity at 1398 eV ("*" feature), as visible in Figure S5a. On the contrary, when the polarization vector is oriented along the main molecular axis the weight of the features changes (*cf.* Figure S5b). In the case of an incidence angle of 60° with respect to the molecular axis, a spectrum similar to the one in Figure S5c is obtained. A full multiX documentation can be found at the link: http://multiplets.web.psi.ch/

# 5. Additional XAS, XLD and XMCD of multilayer and powder samples

**Figure S6** reports the XAS and XLD spectra of 2 MLs and 4 MLs K[Er(COT)$_2$]/Ag(100) and Cp*ErCOT/Ag(100), respectively, while **Figures S7** and **S8** show the circularly polarized and the XMCD spectra. The XMCD values are reported together with the subML samples in **Table 2** of the main text, for direct comparison. **Figure S9** displays the XAS and XMCD spectra of polycrystalline powder of [K(18-c-6)][Er(COT)$_2$]·2THF.

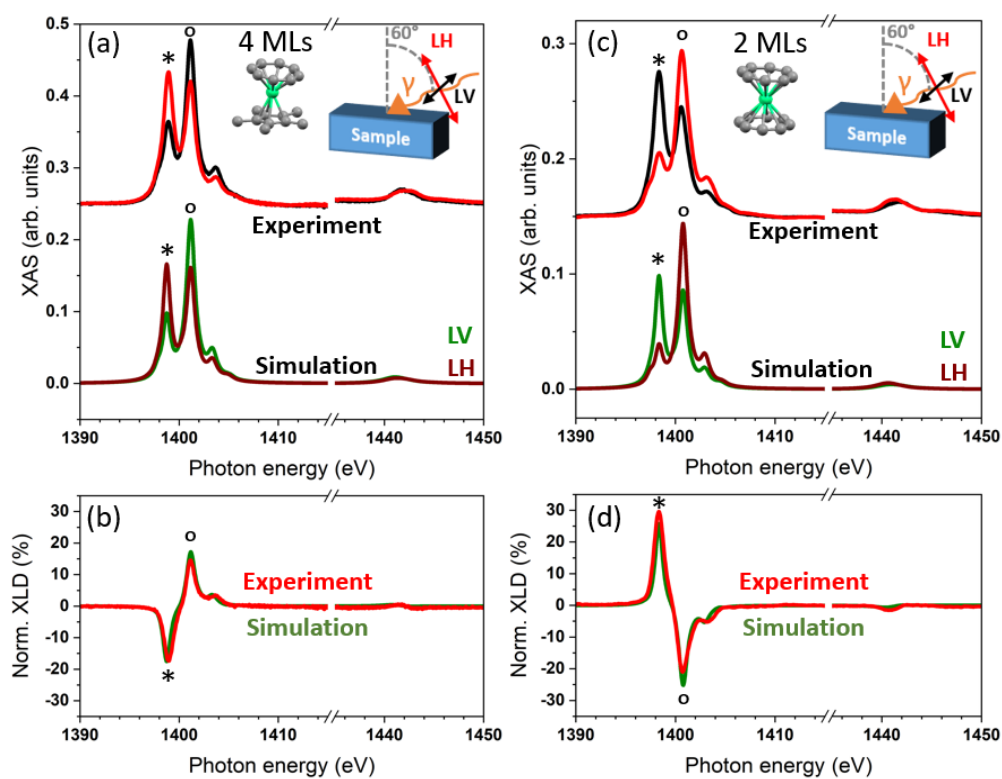

**Figure S6**. Linearly polarized XAS and XLD at the Er M$_{4,5}$-edges measured at 3 K at a grazing angle of 60° from the surface normal direction and multiX-simulated spectra of (a,b) Cp*ErCOT(4 ML)/Ag(100) and (c,d) K[Er(COT)$_2$](2 ML)/Ag(100).

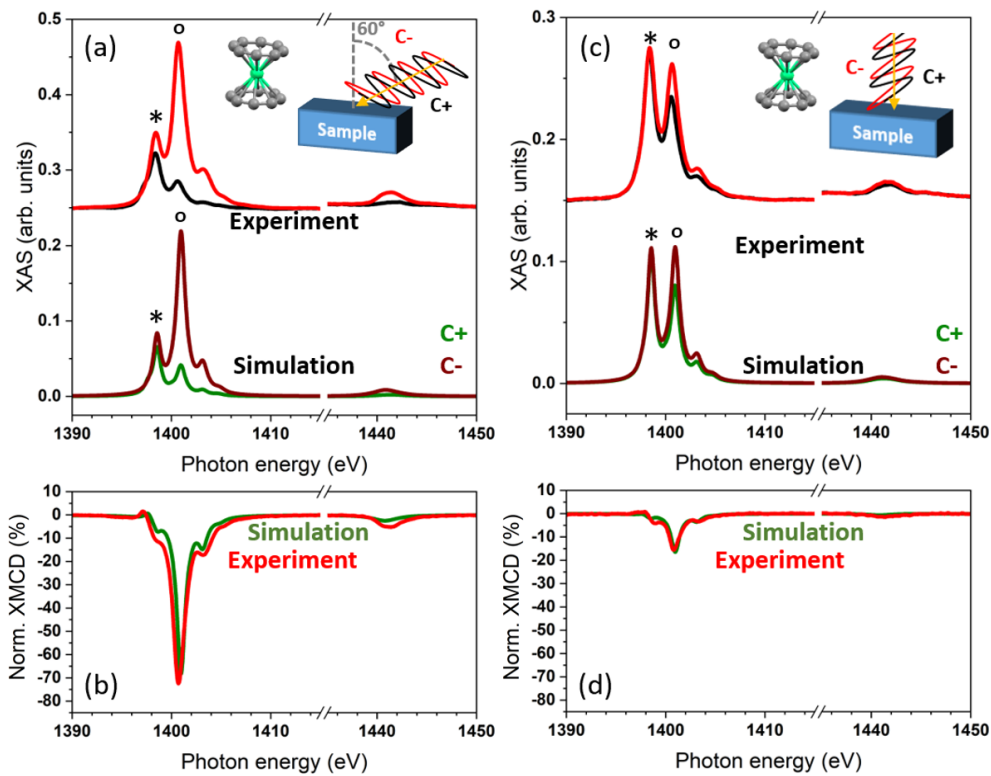

**Figure S7**. Circularly polarized XAS, XMCD and multiX-simulated spectra at the Er $M_{4,5}$-edges recorded in (a,b) grazing and (c,d) normal incidence of coverage of K[Er(COT)$_2$](2 ML)/Ag(100). $T$ = 3 K and $B$ = 6.8 T.

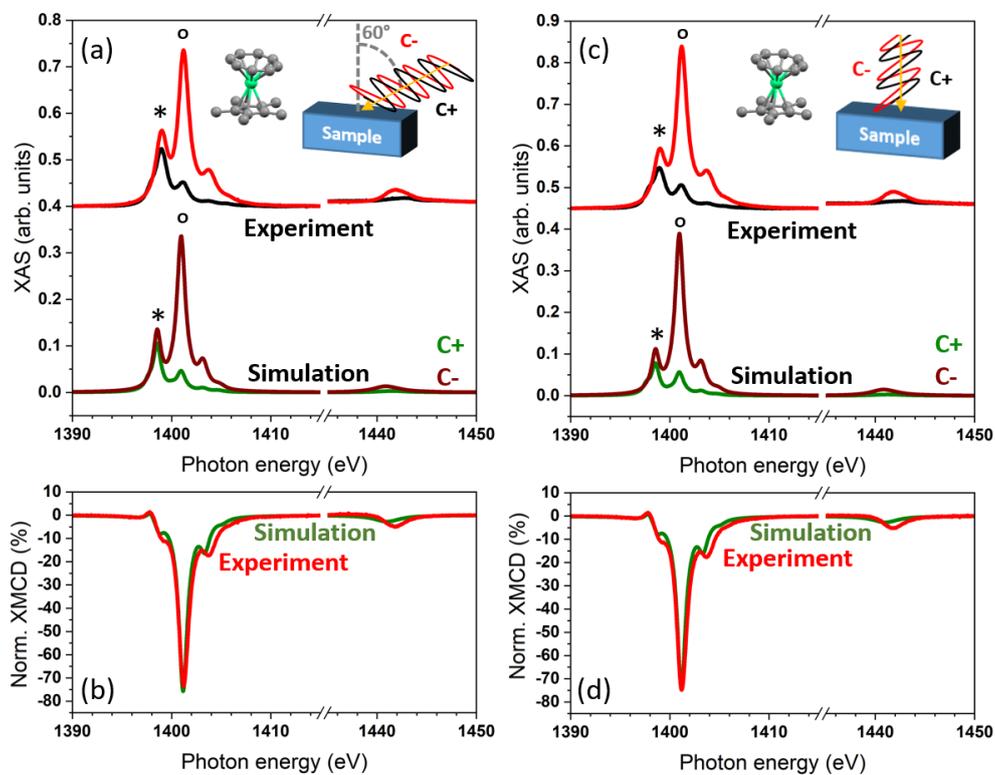

**Figure S8**. Circularly polarized XAS, XMCD recorded at the Er $M_{4,5}$-edges on Cp*ErCOT(4 ML)/Ag(100) and multiX-simulated spectra in (a,b) grazing and (c,d) normal incidence. $T$ = 3 K and $B$ = 6.8 T.

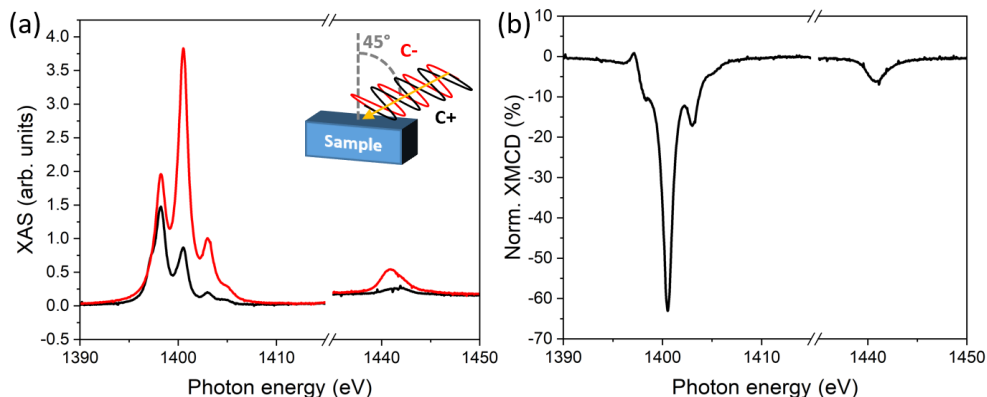

**Figure S9**. (a) Circularly polarized XAS and (b) XMCD recorded at the Er $M_{4,5}$-edges on a polycrystalline powder sample of [K(18-c-6)][Er(COT)$_2$]·2THF. $T$ = 3 K and $B$ = 6.8 T. The powder was pressed in indium foil, which was attached to the sample holder. The preparation and the transfer to the vacuum system of the X-Treme beam line were performed under helium gas atmosphere.

# 6. Extracted magnetic moments from XMCD sum rules

Sum rule analysis[25,26] of the Er $M_{4,5}$ XMCD spectra was performed to obtain the element-specific magnetic moment of erbium. To compare the experimental magnetic moment values to the theoretical ones, sum rule analysis was also performed on the multiX-simulated spectra. The results are reported in **Tables S2** and **S3**. The $<L_Z>$ and $<S_{Zeff}>$ values extracted from the spectra in Figures 4 and 5 are similar to the ones extracted from the corresponding multiX simulations. In order to determine the $<S_Z>$ values from $<S_{Zeff}>$ we assumed that the magnetic dipole moment $<T_Z>$ is proportional to $<S_{Zeff}>$, so when the molecular easy-axis is parallel to the field and the incoming photons the magnitude has the maximum value of the free Er$^{3+}$ ion $<T_Z>$.[27] For other geometries, we scaled the value by the same fraction factor that $<S_{Zeff}>$ scales with respect to its maximum value.

For Cp*ErCOT(0.5 ML)/Ag(100) the total magnetic moment in normal incidence is about half of the expected value for the pristine molecule since only half of the molecules have the magnetic easy-axis aligned with the field. The values in normal incidence are well reproduced by the simulations, but the experimental moments in grazing incidence are smaller than the values obtained by multiX. While the magnetic moment in grazing incidence is expected to be smaller than in normal incidence because of the angle-dependent projection along the beam direction, an anisotropic ordering of the molecules in the substrate plane can reduce the easy-axis contribution in the XMCD spectrum. On the other side, defects in the form of standing-up molecules, as seen in the STM images, can also contribute to a larger out-of-plane magnetization. Since changing the point charges of the carbon atoms or the standing-up vs. lying-down ratio of the molecules worsens the agreement of the simulations with the experimental spectra, we assume other parameters are relevant for the simulation (for example the anisotropic orientation of the lying-down molecules).

For K[Er(COT)$_2$](0.5 ML)/Ag(100) the total magnetic moments of 4.5 ± 0.8 $\mu_B$ in grazing against the 1.9 ± 0.8 $\mu_B$ in normal incidence fits the anisotropy of the net magnetization resulting from the lying-down geometry. The experimental and simulated values are in excellent agreement, corroborating the strength of the point-charge model based on the D$_{8h}$ symmetry with the point charge of the carbon atoms of $q_C$ = 0.25 $e$ and a fraction of 13% of standing-up complexes.

The sum rule results of the multilayer samples show an excellent agreement between the experimental values and the ones extracted from the simulated spectra, as shown in Table S3. At low temperature and high magnetic field, the total Er magnetic moment of the powder sample of the starting material [K(18-c-6)][Er(COT)$_2$]·2THF reported in Figure S9 and Table S4 is about half of the value expected for free, isotropic Er$^{3+}$ ions (9 $\mu_B$). The reduction of ~0.5 compared to the free ion value is consistent with the randomly oriented sample of strongly anisotropic complexes.[28]

**Table S2**. Orbital, spin and total magnetic moment values extracted from the experimental and simulated spectra of samples with 0.5 ML coverages (Figures 4 and 5). The units are [$\mu_B$].

| Cp*ErCOT | Normal | | Grazing | |
|---|---|---|---|---|
| | Experiment | MultiX | Experiment | MultiX |
| $m_L$ | 3.2 ± 0.8 | 3.3 | 2.4 ± 0.6 | 3.2 |
| $m_S$ | 0.9 ± 0.2 | 0.8 | 0.4 ± 0.1 | 0.8 |
| $m_B$ | 5.0 ± 1.0 | 4.9 | 3.3 ± 0.7 | 4.7 |
| K[Er(COT)$_2$] | Normal | | Grazing | |
| | Experiment | MultiX | Experiment | MultiX |
| $m_L$ | 1.2 ± 0.5 | 1.2 | 3.1 ± 0.7 | 3.4 |
| $m_S$ | 0.4 ± 0.3 | 0.3 | 0.7 ± 0.1 | 0.8 |
| $m_B$ | 1.9 ± 0.8 | 1.8 | 4.5 ± 0.8 | 5.0 |

**Table S3**. Orbital, spin and total magnetic moment values extracted from the experimental and simulated spectra of multilayer samples reported in Figures S7 and S8. The units are [$\mu_B$].

| 4 ML Cp*ErCOT | Normal | | Grazing | |
|---|---|---|---|---|
| | Experiment | MultiX | Experiment | MultiX |
| $m_L$ | 3.3 ± 0.8 | 3.3 | 2.9 ± 0.7 | 3.2 |
| $m_S$ | 0.8 ± 0.2 | 0.8 | 0.7 ± 0.1 | 0.8 |
| $m_{tot}$ | 4.9 ± 0.9 | 4.9 | 4.3 ± 0.9 | 4.7 |
| 2 ML K[Er(COT)$_2$] | Normal | | Grazing | |
| | Experiment | MultiX | Experiment | MultiX |
| $m_L$ | 0.5 ± 0.2 | 0.5 | 3.4 ± 0.5 | 3.0 |
| $m_S$ | 0.11 ± 0.03 | 0.1 | 0.7 ± 0.1 | 0.7 |
| $m_{tot}$ | 0.7 ± 0.3 | 0.8 | 4.9 ± 0.6 | 4.4 |

**Table S4**. Orbital, spin and total magnetic moment values extracted from the experimental XAS/XMCD of the polycrystalline powder of [K(18-c-6)][Er(COT)$_2$]·2THF reported in Figure S9. The units are [$\mu_B$].

| | Magnetic Moment |
|---|---|
| $m_L$ | 2.9 ± 0.3 |
| $m_S$ | 0.7 ± 0.1 |
| $m_{tot}$ | 4.3 ± 0.5 |

# 7. Additional hysteresis loops

**Figures S10** and **S11** reports the XMCD-detected magnetic hysteresis loops of Cp*ErCOT/Ag(100) and K[Er(COT)$_2$/Ag(100)], respectively. The spectra are obtained at different coverages and are normalized to the value of saturation of the magnetization. The saturation values of the lowest and largest coverages are reported in Tables S2 and S3.

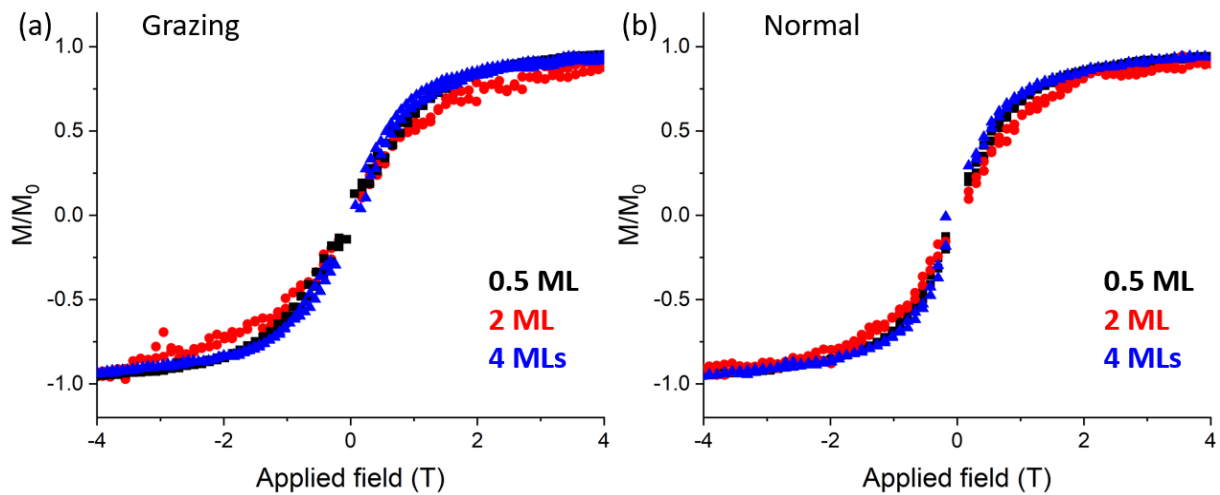

**Figure S10**. XMCD-detected magnetic hysteresis loops of Cp*ErCOT/Ag(100) recorded at 3 K and a rate of 2 T/min on samples with 0.5, 2 and 4 MLs coverage. Normal stands for out-of-plane direction, while grazing (60°) is mostly in-plane.

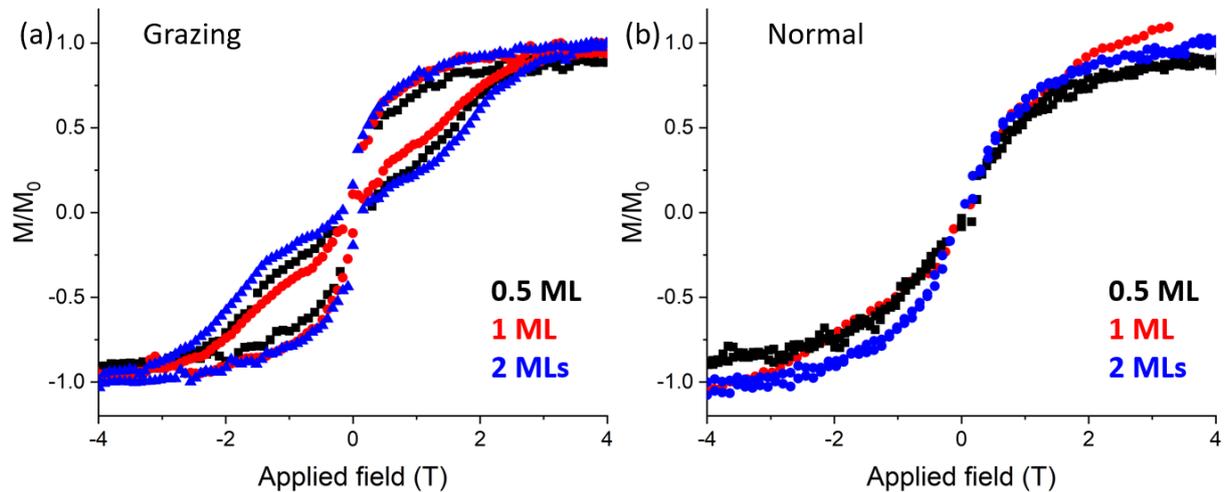

**Figure S11**. XMCD-detected magnetic hysteresis loops of K[Er(COT)$_2$]/Ag(100) recorded at 3 K and a rate of 2 T/min on samples with 0.5, 1 and 2 MLs coverage. Normal stands for out-of-plane direction, while grazing (60°) is mostly in-plane.

## 8. MultiX input file

```
new XMCD
atom Er
ground_state 3d10 4f12

scaler_coulomb 0.85
scaler_so_coupling 0.95
scaler_xtal_field 1.182
threshold_corr 22.5

core_hole_broad 0.50
deltag1         1.00
wming1 1398.0
wmaxg1 1440.0
temperature 3 #used for the population of the sub-states

#the beam direction is in the xyz frame, where xy is the substrate
#plane (y pointing "up" in the experimental configuration) and z is
#out-of-plane direction

bfield 6.8
beam_in    -0.886 -0.5 0   #beam direction in grazing 60 deg
bfield_dir -0.886 -0.5 0 #magnetic field direction in grazing 60 deg

begin_xtal
1.000000 [Angstroems] x y z q radius
 -1.8310    1.9120    0        0.2500
 -1.2947    1.9120   -1.2947   0.2500
  0         1.9120   -1.8310   0.2500
  1.2947    1.9120   -1.2947   0.2500
  1.8310    1.9120    0        0.2500
  1.2947    1.9120    1.2947   0.2500
  0         1.9120    1.8310   0.2500
 -1.2947    1.9120    1.2947   0.2500
 -1.8310   -1.9120    0        0.2500
 -1.2947   -1.9120   -1.2947   0.2500
  0        -1.9120   -1.8310   0.2500
  1.2947   -1.9120   -1.2947   0.2500
  1.8310   -1.9120    0        0.2500
  1.2947   -1.9120    1.2947   0.2500
  0        -1.9120    1.8310   0.2500
 -1.2947   -1.9120    1.2947   0.2500
end_xtal

spect_emin   1390
spect_emax   1450

emid_xmcd    1420
QNprint      14
GSanalyze
```